\documentclass[twocolumn]{aastex7}

\usepackage{xspace}
\usepackage{amsmath}

\newcommand{\Msol}{\ensuremath{\mathrm{M}_{\odot}}\xspace}

\newcommand{\E}{$E\left(B-V\right)$}

\newcommand{\R}{$f_\mathrm{em}/f_\mathrm{abs}$}

\newcommand{\ccm}{\ensuremath{\mathrm{cm}^{-3}}\xspace}
\newcommand{\kms}{\ensuremath{\mathrm{km\,s^{-1}}}\xspace}
\newcommand{\mdot}{\ensuremath{\Msol\,\mathrm{yr}^{-1}}\xspace}

\usepackage{xcolor}

\definecolor{darkorange}{HTML}{A34E18}

\begin{document}

\include{response}
\clearpage

\title{Little Red Dots as Shock-Active High-Energy Neutrino Sources}

\newcommand{\PSI}{\affiliation{Planetary Science Institute, 1700 East Fort
  Lowell Road, Suite 106,Tucson, AZ 85719-2395 USA}}
\newcommand{\HS}{\affiliation{Hamburger Sternwarte, Gojenbergsweg 112, 21029 Hamburg, Germany}}
\newcommand{\IFA}{\affiliation{Institute for Astronomy, University of Hawai’i at Manoa, 2680 Woodlawn Dr., Honolulu, HI 96822, USA}}
\newcommand{\VT}{\affiliation{Department of Physics, Virginia Tech,
    850 West Campus  Drive, Blacksburg VA, 24061, USA}}
\newcommand{\GRFP}{\altaffiliation{National Science Foundation Graduate Research Fellow}}

\newcommand{\FINESST}{\altaffiliation{NASA FINESST Future Investigator}}
\newcommand{\NHFPE}{\altaffiliation{NHFP Einstein Fellow}}
\newcommand{\UIUC}{\affiliation{Department of Astronomy, University of Illinois Urbana-Champaign, 1002 West Green Street, Urbana, IL 61801, USA}}
\newcommand{\NSFSIMS}{\affiliation{NSF-Simons AI Institute for the Sky (SkAI), 172 E. Chestnut St., Chicago, IL 60611, USA}}

\newcommand{\STSci}{\affiliation{Space Telescope Science Institute, 3700 San Martin Drive, Baltimore, MD 21218-2410, USA}}
\newcommand{\FSU}{\affiliation{Department of Physics, Florida State
    University, Tallahassee, FL 32306, USA}}
\newcommand{\Carnegie}{\affiliation{Observatories of the Carnegie
    Institution for Science, 813 Santa Barbara St., Pasadena, CA 91101, USA}}
\newcommand{\MSU}{\affiliation{Department of Physics \& Astronomy,
    Michigan State University, East Lansing, MI, USA}}
\newcommand{\TAMU}{\affiliation{George P. and Cynthia Woods Mitchell
    Institute for Fundamental Physics and Astronomy,
    Department of Physics and Astronomy, Texas 
             A\&M University, College Station, TX 77843, USA}}
\newcommand{\IALP}{\affiliation{Instituto de Astrof\'isica de La Plata
    (IALP), CONICET, Paseo del Bosque S/N, B1900FWA La Plata, Argentina}}
\newcommand{\LaPlata}{\affiliation{Facultad de Ciencias Astron\'omicas
    y Geof\'isicas Universidad Nacional de La Plata, Paseo del Bosque,
    B1900FWA, La Plata, Argentina}}
\newcommand{\WPI}{\affiliation{Kavli Institute for the Physics and
    Mathematics of the Universe (WPI), The University of Tokyo,
    Kashiwa, 277-8583 Chiba, Japan}} 

\newcommand{\ICE}{\affiliation{Institute of Space Sciences (ICE,
    CSIC), Campus UAB, Carrer de Can Magrans, s/n, E-08193 Barcelona, Spain}}

\newcommand{\IEEC}{\affiliation{Institut d’Estudis Espacials de
    Catalunya (IEEC), E-08034  Barcelona, Spain}} 

\newcommand{\LCO}{\affiliation{Las Campanas Observatory, Carnegie
    Observatories, Casilla 601, La Serena, Chile}} 

\newcommand{\Aarhus}{\affiliation{Department of Physics and Astronomy,
    Aarhus University, Ny  Munkegade 120, DK-8000 Aarhus C, Denmark.}} 

\newcommand{\OU}{\affiliation{Homer L.~Dodge Department of Physics and
  Astronomy, University of Oklahoma, 440 W. Brooks, Rm 100, Norman, OK
  73019-2061}}  

\newcommand{\UCSC}{\affiliation{Department of Astronomy and Astrophysics,
  University of California, Santa Cruz, CA 95064, USA}} 
\newcommand{\Melbourne}{\affiliation{School of Physics, The University of
  Melbourne, VIC 3010, Australia}}

\newcommand{\LPNHE}{\affiliation{LPNHE, (CNRS/IN2P3, Sorbonne
  Universit\'e, Universit\'e Paris Cit\'e), Laboratoire de Physique
  Nucl\'eaire et de Hautes \'Energies, 75005, Paris, France}}

\newcommand{\Princeton}{\affiliation{Princeton University, 4 Ivy Lane,
    Princeton, NJ 08544, USA}}

\newcommand{\Berkeley}{\affiliation{Department of Astronomy,
    University of California, Berkeley, CA 94720-3411, USA}}

\newcommand{\Tsinghua}{\affiliation{Physics Department, Tsinghua
    University, Beijing, 100084, China}}

\newcommand{\Thailand}{\affiliation{National Astronomical Research
    Institute of Thailand, 260 Moo 4, Donkaew, Maerim, Chiang Mai,
    50180, Thailand}}

\newcommand{\UVA}{\affiliation{Department of Astronomy, University of
    Virginia, 530 McCormick Rd, Charlottesville, VA 22904, USA}}

\newcommand{\LJMU}{\affiliation{Astrophysics Research Institute,
    Liverpool John Moores University, 146 Brownlow Hill, Liverpool L3
    5RF, UK}}

\newcommand{\MPIA}{\affiliation{Max-Planck-Institut f\"ur Astrophysik,
    Karl-Schwarzschild Stra{\ss}e 1, 85748 Garching, Germany}}

\newcommand{\JHU}{\affiliation{Physics and Astronomy Department,
    Johns Hopkins University, Baltimore, MD 21218, USA}}

\newcommand{\OSU}{\affiliation{Department of Astronomy, The Ohio State
    University, Columbus, OH, USA}}

\newcommand{\CCAP}{\affiliation{Center for Cosmology and Astroparticle
    Physics, The Ohio State University, Columbus, OH, USA}}

\newcommand{\MIT}{\affiliation{Department of Physics and Kavli Institute for Astrophysics and Space Research, Massachusetts Institute of Technology, 77 Massachusetts Avenue, Cambridge, MA 02139, USA}}

\newcommand{\IST}{
\affiliation{Department of Physics,
Institute of Science Tokyo,
2-12-1 Ookayama,
Meguro-ku, Tokyo 152-8551, Japan}
}

\newcommand{\KIPMU}{
\affiliation{Kavli Institute for the Physics and Mathematics of the Universe (Kavli IPMU, WPI),
UTIAS,
The University of Tokyo,
Kashiwa, Chiba 277-8583, Japan}
}

\newcommand{\nextinstitute}{\affiliation{Put the institute of the new author here}}

\author[0000-0001-5888-2542]{T.~Mera}
\email{tycomera@gmail.com}
\IFA

\author[0000-0002-5221-7557]{C. Ashall}
\email{cashall@hawaii.edu}
\IFA

\author[0000-0001-6142-6556]{S. Horiuchi }
\email{horiuchi@vt.edu}
\IST
\VT
\KIPMU

\author[0000-0003-3997-5705]{R. P. Naidu}
\email{rnaidu@hawaii.edu}
\IFA

\author[0000-0001-7186-105X]{K. Medler}
\email{kyle.medler@sky.com}
\IFA

\author[0000-0002-4338-6586]{P.~Hoeflich}
\email{phoeflich77@gmail.com}
\FSU

\submitjournal{ApJ}

\received{\today}
\revised{\today}
\accepted{\today}

\begin{abstract}
Little Red Dots (LRDs) are compact, high-redshift
sources whose physical nature remains uncertain. Their optical spectra motivate a scenario in which their
emission may be powered by shocks interacting with dense surrounding
material. We investigate whether such interactions can power LRDs and contribute
to the diffuse high-energy neutrino intensity measured by IceCube. In our simplified model, a fast
central-engine outflow drives a shock through stationary or slow
outflowing material, followed by a prescribed quasi-stationary interaction
near the LRD photosphere. We consider the time-averaged neutrino
emission from shock-active LRDs without requiring a single shock to
persist throughout the LRD lifetime. We explore parameter ranges motivated by
the observed and inferred properties
of LRDs, finding solutions with shock luminosities
$2.0\times10^{43}\lesssim L_s\lesssim3.5\times10^{44}$~erg~s$^{-1}$.
Using an analytical framework for cosmic-ray acceleration and hadronic
interactions in dense shock environments, we calculate the resulting
high-energy neutrino emission and integrate it over the cosmological
LRD population. For our fiducial cosmic-ray parameters, the average
predicted contribution below $2\times10^5$~GeV increases from
$\sim0.2\%$ for the lowest-luminosity quintile to $\sim2\%$ for the
highest-luminosity quintile, while an illustrative higher-efficiency
case reaches $\sim13\%$ of the IceCube diffuse neutrino intensity.
These estimates assume all observed LRDs are shock-active. A smaller
active fraction or a reduced covering factor would decrease the predicted intensity, whereas survey
incompleteness could increase it. Improved models and population
synthesis are needed to constrain the total LRD contribution.

\end{abstract}


\section{Introduction} \label{sec:Intro}

The basic nature of Little Red Dots (LRDs; \citealt{2024Matthee}) remains debated, with
interpretations ranging from accreting supermassive black holes to seed black holes to
supermassive stars
\citep{2025Nandal,2025deGraaffa,2026Naidu_nature,2026Juodvzbalis,2026Madau,
2026Greene,2026Chisholm,2026Martins,2026Zwick,2026Santarelli}.
It has recently been shown that the observed properties of LRDs bear
many resemblances to those of Type~IIn supernovae (SNe~IIn)
\citep{2007Izotov,2008Izotov,2026Naidu,2026Ashall}. In particular,
both exhibit similar optical spectral morphologies characterized by
strong Balmer emission, broad line wings, and, in some cases, P-Cygni
absorption. Moreover, LRDs and SNe~IIn occupy similar effective
temperature regimes, which, when combined with these spectral
similarities, may point toward a common physical configuration in
which fast outflows interact with a dense, slower-moving surrounding
medium, producing strong shocks and an optically thick
pseudo-photosphere
\citep{2003Hamuy,2004Deng,2007Prieto,2015Fox,2023Uno}.

Such dense shock-interaction environments are also favorable sites for
high-energy neutrino production. In SNe~IIn, shocks can accelerate
cosmic-ray protons that interact with the dense surrounding material,
and these systems have been shown theoretically to contribute to the
diffuse high-energy neutrino background
\citep{2018Murase,2024Murase,2026Wasserman}. LRDs have also recently
been proposed as potential high-energy neutrino sources through
interacting bipolar outflows from central black hole engines
\citep{2026Kuze}. Motivated by this background, we instead investigate whether shock interactions
in LRDs can produce high-energy neutrinos and contribute to the diffuse
neutrino background.

In Section~\ref{sec:Models}, we construct
a simplified shock-interaction model for LRDs and outline the analytical framework used to estimate the resulting
high-energy neutrino emission. In Section~\ref{sec:Results}, we explore
the physically motivated parameter space of the model and determine
the potential contribution of LRDs to the diffuse neutrino background.
Finally, we discuss the implications of our results and summarize our
conclusions in Section~\ref{sec:Conclusions}.

\begin{figure*}[t]
    \centering
    \includegraphics[width=0.99\linewidth]{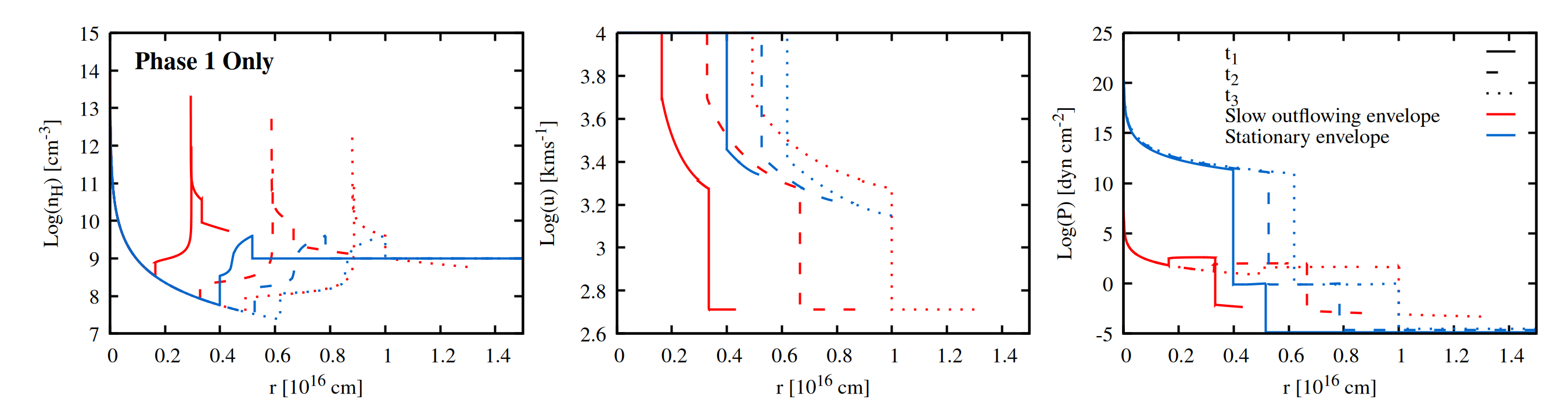}
\caption{Evolution of the initial propagating shock (phase 1) into slow-outflowing (red) and stationary (blue) envelopes, calculated using a modified version of the SPICE software \citep{2016Dragulin}. Phase 2 is not explicitly modeled hydrodynamically. Instead, we prescribe a quasi-stationary shock near the photosphere, with its luminosity determined by the kinetic-energy flux of the interacting material. As an illustrative example, we show two models with $M_{\rm eng}=10^{5}~\Msol$, $n_{\rm H}=10^{9}$ \ccm, $\dot{M}_{\rm eng}=0.3~\mdot$, and $v_{\rm eng}=10^{4}~\kms$, with the photosphere defined at $R_{\rm ph}=10^{16}$~cm. From left to right, we show the hydrogen number density, $n_{\rm H}$ (\ccm), velocity, $u$ (\kms), and pressure, $P$ (dyn cm$^{-2}$), as functions of radius. The three line styles correspond to $t_{1,2,3}=0.45$, $0.90$, and $1.36$~yr for the slow-outflowing envelope and $0.34$, $0.67$, and $1.01$~yr for the stationary envelope. Other models with similarly structured envelopes will exhibit the same qualitative shock structure but with different physical and temporal scalings. On average, the outflowing models reach $R_{\rm ph}$ after 2.18~yr, while the stationary models reach it after 2.15~yr.}
    \label{fig:shock}
\end{figure*}

\section{LRD Model} \label{sec:Models}

We consider LRDs in which a powerful central engine drives fast outflows into a dense, slower-moving medium. Such conditions may arise in gas-enshrouded accreting black holes, broadly termed black-hole stars (BH*s). Here, BH*s encompass configurations in which a black hole is embedded within a dense gaseous envelope, with classical quasi-stars representing only one possible realization \citep{2008Begelman,2025Kido,2025deGraaffb,2026Begelman,2026Santarelli,2026Santarellib,2026Cantiello,2026Hassan,2026RomanGarza,2026Gentile,2026Naidu_nature,2026Sun,2026Sunb,2026Rusakov,2026Kokorev,2026Asada,2026Matthee}. The dense surrounding material may originate from pre-collapse supermassive star (SMS) mass loss \citep{2026Nandalb,2026Torralbab}, BH-seed formation through rapid accretion or stellar collisions \citep{2026Nandala,2025Vergara,2026Rantala,2026Dekel}, or dense environments surrounding newly formed direct-collapse black holes (DCBHs) \citep{2024Kiyuna,2026Baggen,2026Hazlett,2026Chon,2026Pacucci,2026Whalen}. Other mechanisms for supplying dense gas around black holes include accretion in gas-rich nuclear star clusters \citep{2014Alexander} or the formation of optically thick envelopes from stellar tidal-disruption debris \citep{1997LoebUlmer}. Super-Eddington accretion may also generate the required velocity structure, producing fast outflows alongside slower winds and fallback material in fountain-flow configurations \citep{2025Liu,2026LiuHo,2026Kaaz}. Modern SMS evolution challenges the original quasi-star formation pathway \citep{2013Hosokawa,2016Umeda,2017Woods,2018Haemmerle,2023Herrington}, but recent studies have begun exploring alternative mechanisms for forming and sustaining dense gaseous environments after collapse \citep{2026RomanGarza,2026Nandalb,2026Torralbab,2026Chon}. These scenarios provide possible environments for fast--slow interactions, although not all necessarily produce strong shocks. Our model therefore applies specifically to systems in which such interactions develop.

We adopt spherical symmetry as our fiducial approximation ($f_\Omega=1$), while exploring smaller covering factors to represent incomplete angular overlap between the outflow and surrounding gas. Such asymmetries may arise from the winds, inflows, and fallback expected in LRD environments \citep{2026Kaaz,2026Matthee,2026LiuHo,2026Sunb}. Motivated by a characteristic photospheric radius of $R_{\rm ph}\sim10^{16}$~cm \citep{2025deGraaffb,2025Kido,2025Liu,2026Naidua,2026Umeda,2026Sunb}, we model an initially propagating shock (phase 1) followed by a prescribed quasi-stationary interaction near $R_{\rm ph}$ (phase 2). We adopt an illustrative interaction duration of $\sim40$~yr, motivated by the inferred rest-frame variability timescale of a gravitationally lensed LRD \citep{2025Zhang,2026Zhang}. Similar timescales may arise in super-Eddington fountain flows, for which dynamical timescales of $\sim100$~yr have been estimated \citep{2026Kaaz}. Although these estimates do not directly constrain shock lifetimes, they motivate considering interactions much shorter than the inferred $\sim10$--$20$~Myr LRD lifetime \citep{2026Sun,2026Santarelli,2026RomanGarza}. Sustaining a single stalled shock over that lifetime remains uncertain because radiative cooling can drive shock oscillations and shell instabilities \citep{1982ChevalierImamura,1984Imamura,1988Gaetz,1994Vishniac,2018Steinberg}. Instead, continued outflows and fallback may produce successive shocks, allowing interactions to be episodic or time-dependent \citep{2009vanMarle,2018Steinberg}. Our diffuse neutrino calculation depends on the instantaneous comoving number density of active LRDs, $n_{\rm LRD}(z)$ \citep{2026Kapoor}, and the mean shock luminosity derived from the outflow kinetic power, without requiring a single stalled shock to remain stable over Myr timescales. Therefore, our predictions represent an optimistic scenario in which all observed LRDs are shock-active at a given redshift, while a smaller interacting fraction would proportionally reduce the diffuse contribution. The formation and stability of the prescribed quasi-stationary interaction remain uncertain in our setup.

\subsection{Shock Interactions}\label{sec:shocks}

To model the hydrodynamic interaction, we use a modified version of the
Supernovae Progenitor Interaction Calculator for parameterized
Environments (SPICE; \citealt{2016Dragulin}). SPICE is a semi-analytic,
spherically symmetric framework for wind--environment interactions,
building upon analytic and self-similar shock solutions
\citep{1963Parker,1977Weaver,1982Chevalier,1983Chevalier}. The
hydrodynamic equations and Rankine--Hugoniot jump conditions determine
the density, velocity, and pressure across the reverse shock, contact
discontinuity, and forward shock. SPICE considers both constant-density
($p=0$) and wind-like ($p=2$) environments and uses scale-free relations
based on the Buckingham $\Pi$ theorem
\citep{1914Buckingham,1959Sedov}, allowing solutions to be efficiently
rescaled over a broad parameter space. This makes SPICE applicable to general spherically symmetric
wind--environment interactions. Here, we modify SPICE for the densities,
luminosities, and central-engine masses relevant to LRDs.

\subsubsection{Modifications to SPICE}

A wind-like density profile, $\rho_{\rm CSM}\propto r^{-2}$ ($p=2$),
is commonly adopted for steady
mass-loss events, while $p=0$ is used for stationary material
\citep{1994FranssonA,2003Chevalier,2014Fransson}. To encompass the different scenarios described above we use
both $p=0$ and $p=2$ throughout this work. The corresponding SPICE model
is specified by the interaction time, $t$, the mass-loss rate and
velocity, $\dot{M}_{\rm csm}$ and $v_{\rm csm}$, and the central-engine
mass-loss rate and velocity, $\dot{M}_{\rm eng}$ and $v_{\rm eng}$.
For the stationary case ($p=0$), the envelope is instead specified
by its density, with $v_{\rm csm}=0$.

Motivated by the physical pictures outlined above, we assume that shock-active LRDs are powered by an initial outward-propagating shock followed by a prescribed quasi-stationary interaction
near the photospheric radius, $R_{\rm ph}$, with
characteristic hydrogen number density $n_{\rm H}$ and central-engine
mass $M_{\rm eng}$. Rather than directly sampling $t$,
$\dot{M}_{\rm csm}$, and $v_{\rm csm}$, we parameterize the model by
$M_{\rm eng}$, $\dot{M}_{\rm eng}$, $v_{\rm eng}$, and particle density, $n_{\rm H}$,
and use these quantities to determine the initial SPICE solutions.

Phase~1 describes
the initial expansion of the shock to $R_{\rm ph}$.
For the outflowing case ($p=2$), we set
$v_{\rm csm}$ equal to the escape velocity at $R_{\rm ph}$ for a given
$M_{\rm eng}$. We determine $\dot{M}_{\rm csm}$ by requiring the CSM
density at $R_{\rm ph}$ to equal the adopted $n_{\rm H}$, giving
\begin{equation} \label{eq:density}
    \rho_{\rm csm}(R_{\rm ph})
    =
    \frac{\dot{M}_{\rm csm}}
    {4\pi R_{\rm ph}^{2}v_{\rm csm}}
    =
    m_p n_{\rm H},
\end{equation}
where $m_p$ is the proton mass.
For the stationary case ($p=0$), we instead adopt
$\rho_{\rm csm}=m_p n_{\rm H}$ throughout the CSM.
An example of this initial evolution is shown in
Figure~\ref{fig:shock}. The kinetic luminosity processed by the
forward shock during this phase is
\begin{equation}
    L_s =
    2\pi f_{\Omega}\rho_{\rm csm}V_s^3R_s^2,
\end{equation}
where $f_{\Omega}$ is the CSM covering factor and $V_s$ and $R_s$ are
the forward-shock velocity and radius.

During Phase~2, we assume that the shock remains stalled at
$R_{\rm ph}$. For the outflowing case ($p=2$), the CSM velocity remains fixed, while
$\dot{M}_{\rm csm}$ is set by the mass-loss rate required to maintain
$n_{\rm H}$ at $R_{\rm ph}$. When
$\dot{M}_{\rm eng}>\dot{M}_{\rm csm}$, the excess material falls back
toward the central engine at a rate
$\dot{M}_{\rm fb}=\dot{M}_{\rm eng}-\dot{M}_{\rm csm}$.
Assuming this material impacts the stalled shock with a speed
$v_{\rm csm}$, the total kinetic luminosity processed by the shock is
\begin{equation}
    L_s =
    \frac{1}{2}f_{\Omega}
    \left(
    \dot{M}_{\rm eng}v_{\rm eng}^{2}
    +
    \dot{M}_{\rm fb}v_{\rm csm}^{2}
    \right).
\end{equation}
For the stationary case ($p=0$), only the first term of the equation contributes and we assume no mass escapes ($\dot{M}_{\rm csm}=0$).
Within our parameterized model, this stalled solution can, in principle,
be maintained indefinitely, provided that the total escaped mass does
not exceed the available central-engine mass.
However, the long-term stability of this configuration remains uncertain.
We therefore limit the combined duration of Phases~1 and 2 to
the illustrative $\sim40$~yr interaction timescale described above.
This represents an individual interaction episode, while continued
outflows and fallback may produce successive shocks over the much longer
$\sim10$--$20$~Myr lifetime of an LRD.

\subsection{Neutrino Emission} \label{sec:neutrino}

The shock luminosity provides the energy reservoir available for
nonthermal particle acceleration and subsequent high-energy emission.
To calculate the resulting neutrino signal, we follow the analytical
framework of \citet{2024Murase}, which models high-energy
emission from shocks interacting with dense circumstellar material. 

The maximum cosmic-ray energy is determined by balancing the proton acceleration time against the relevant loss and escape timescales. The high hydrogen densities in our LRD models make inelastic $pp$ interactions the dominant energy-loss process for accelerated protons. We therefore take the maximum energy to be limited by $pp$ losses or particle escape. Following \citet{2024Murase}, the
$pp$-limited maximum energy is
\begin{equation}
    E_{\rm cr,max}^{pp}
    =
    \frac{3}{20}
    \frac{eBv_{\rm rel}^{2}}{c}
    t_{pp},
\end{equation}
where
$t_{pp}=(\kappa_{pp}\sigma_{pp}n_{\rm H}c)^{-1}$,
$\kappa_{pp}\simeq0.5$, $\sigma_{pp}\simeq3\times10^{-26}$~cm$^2$,
and the magnetic field is parameterized by
$B^2/(8\pi)=3\epsilon_B L_s/(4\pi R_s^2v_{\rm rel})$.
Here, $\epsilon_B$ is the fraction of the shock energy density
carried by magnetic fields. We adopt $\epsilon_B=10^{-2}$ \citep{2012Maeda,2014Caprioli}.
Here, $v_{\rm rel}$ is the upstream velocity measured in the shock
frame. During Phase~1, $v_{\rm rel}$ is determined by the evolving
shock solution, while for the stalled shock in Phase~2 we take
$v_{\rm rel}\simeq v_{\rm eng}$. We additionally consider escape from
the acceleration region,
\begin{equation}
    E_{\rm cr,max}^{\rm esc}
    =
    \frac{3}{20}
    \frac{eBv_{\rm rel}l_{\rm esc}}{c},
\end{equation}
where we take $l_{\rm esc}=R_s$. The maximum cosmic-ray energy is then
$E_{\rm cr,max}=\min(E_{\rm cr,max}^{pp},E_{\rm cr,max}^{\rm esc})$. 

We assume that a fraction $\epsilon_{\rm cr}=0.1$ of the shock luminosity
is converted into accelerated cosmic rays with a power-law spectrum of
index $s_{\rm cr}$. The differential all-flavor neutrino luminosity is then approximated as
\begin{equation} \label{eq:neutrino}
    \varepsilon_\nu L_{\varepsilon_\nu}
    \approx
    \frac{1}{2}
    \min[f_{pp},1]
    \frac{\epsilon_{\rm cr}L_s}{\mathcal{R}_{\rm cr10}}
    \left(
    \frac{\varepsilon_\nu}{0.4~{\rm GeV}}
    \right)^{2-s_{\rm cr}},
\end{equation}
where
$\mathcal{R}_{\rm cr10}\equiv
\epsilon_{\rm cr}L_s/
(\epsilon_p^2 d\dot{N}_{\rm cr}/d\epsilon_p)|_{10\,{\rm GeV}}$
is the cosmic-ray bolometric correction evaluated between
$E_{\rm cr,min}$ and $E_{\rm cr,max}$. We use
$d\dot{N}_{\rm cr}/d\epsilon_p\propto\epsilon_p^{-s_{\rm cr}}$, $s_{\rm cr}=2.1$, $E_{\rm cr,min}=1$ GeV, and calculate the $pp$ interaction efficiency as
\begin{equation}
    f_{pp}
    \approx
    \kappa_{pp}\sigma_{pp}
    \left(\frac{\rho_{\rm csm}}{m_{\rm H}}\right)
    R_s\left(\frac{c}{v_{\rm rel}}\right).
\end{equation}

\begin{figure*}[t]
    \centering
    \includegraphics[width=0.99\linewidth]{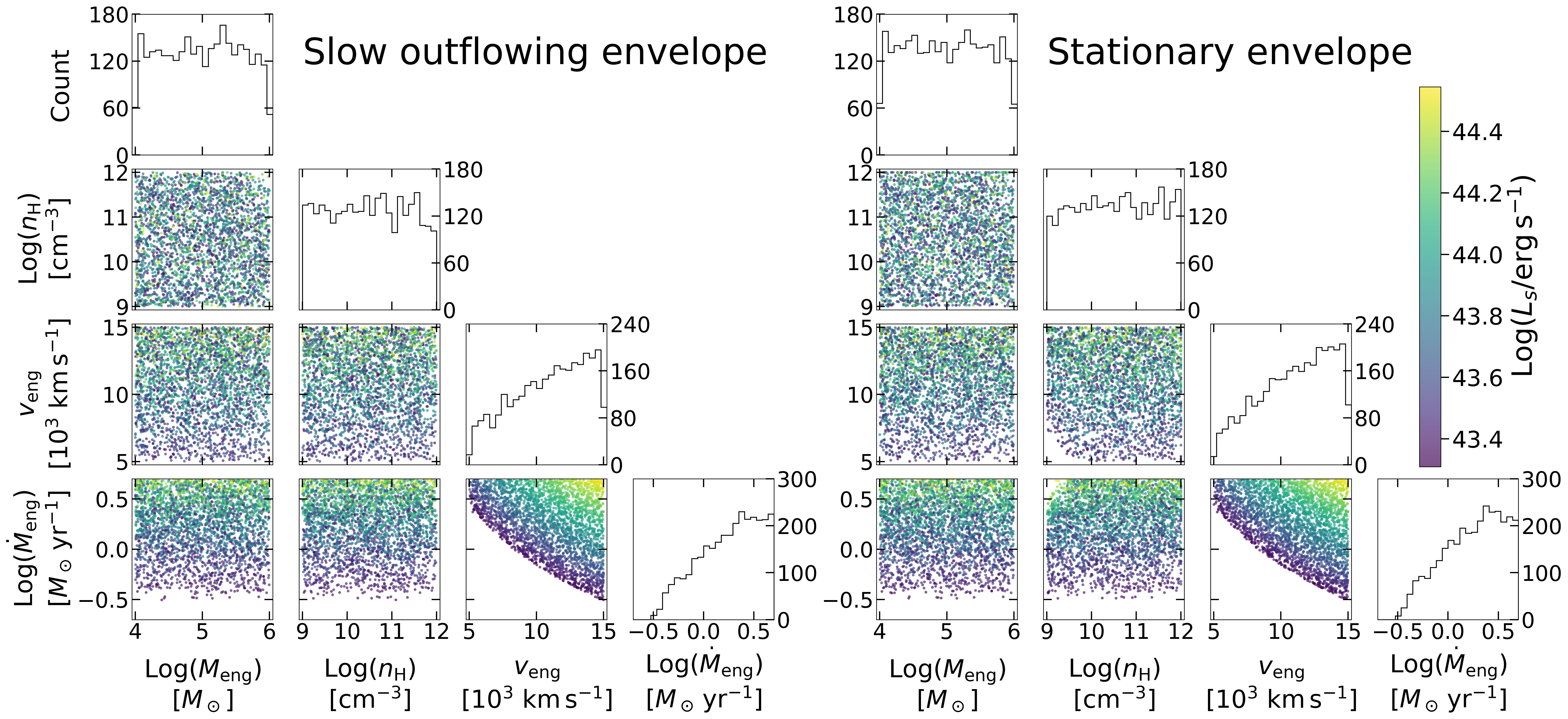}
    \caption{ Corner plot showing viable solutions of the slow outflowing (left) and stationary envelope (right) shock models
constrained by the observed bolometric luminosities of LRDs, the
diffuse all-flavor neutrino intensity measured by IceCube, and the
requirement $M_{\rm esc}\leq M_{\rm eng}$. The color scale indicates
the total shock luminosity, $L_s$, of each model. Note, a decrease in $f_{\Omega}$ would proportionally decrease $L_s$.}
    \label{fig:solutions}
\end{figure*}

\subsubsection{Diffuse Neutrino Contribution}

To determine the total neutrino contribution from the LRD population,
we use the instantaneous comoving number density of active LRDs,
$n_{\rm LRD}(z)$, from \citet{2026Kapoor}:
\begin{equation}
n_{\rm LRD}(z)
=
n_0 f(z)
\exp\left[
-\frac{\left(\ln(1+z)-\mu_z\right)^2}
{2\sigma_z^2}
\right],
\end{equation}
where
$f(z)=(1+z)^{3/2}/[q(1+z)^{1/2}-1]^2$,
$n_0=3\times10^{-6}$~cMpc$^{-3}$,
$\mu_z=\ln(1+z_0)$ with $z_0=5$, $\sigma_z=0.28$, and
$q=0.903$. Note that $n_{\rm LRD}$ may be underestimated due to the low
completeness of current surveys toward faint LRDs \citep{2025Ma,2026Ma,2026Kapoor}. 

The diffuse all-flavor
neutrino intensity is then obtained by integrating the emission from
individual LRDs over their cosmological distribution,
\begin{equation}\label{eq:diffuse}
\begin{split}
E_{\rm obs}^{2}\Phi_{\nu}(E_{\rm obs})
={}&
\frac{c}{4\pi}
\int dz\,
\frac{n_{\rm LRD}(z)}
{H(z)(1+z)^2}
\\
&\times
\left\langle
E_{\rm em}L_{E_{\rm em}}
\left[(1+z)E_{\rm obs}\right]
\right\rangle .
\end{split}
\end{equation}
Here, $E_{\rm em}=(1+z)E_{\rm obs}$, $H(z)$ is the Hubble
expansion rate, with $H_0=67.4~{\rm km~s^{-1}~Mpc^{-1}}$. Since $n_{\rm LRD}(z)$ represents the instantaneous
number density of active sources rather than an event rate, we use the
interaction episode-averaged source spectrum rather than the time-integrated
spectrum of an individual LRD.

\begin{figure}
    \centering
    \includegraphics[width=0.99\linewidth]{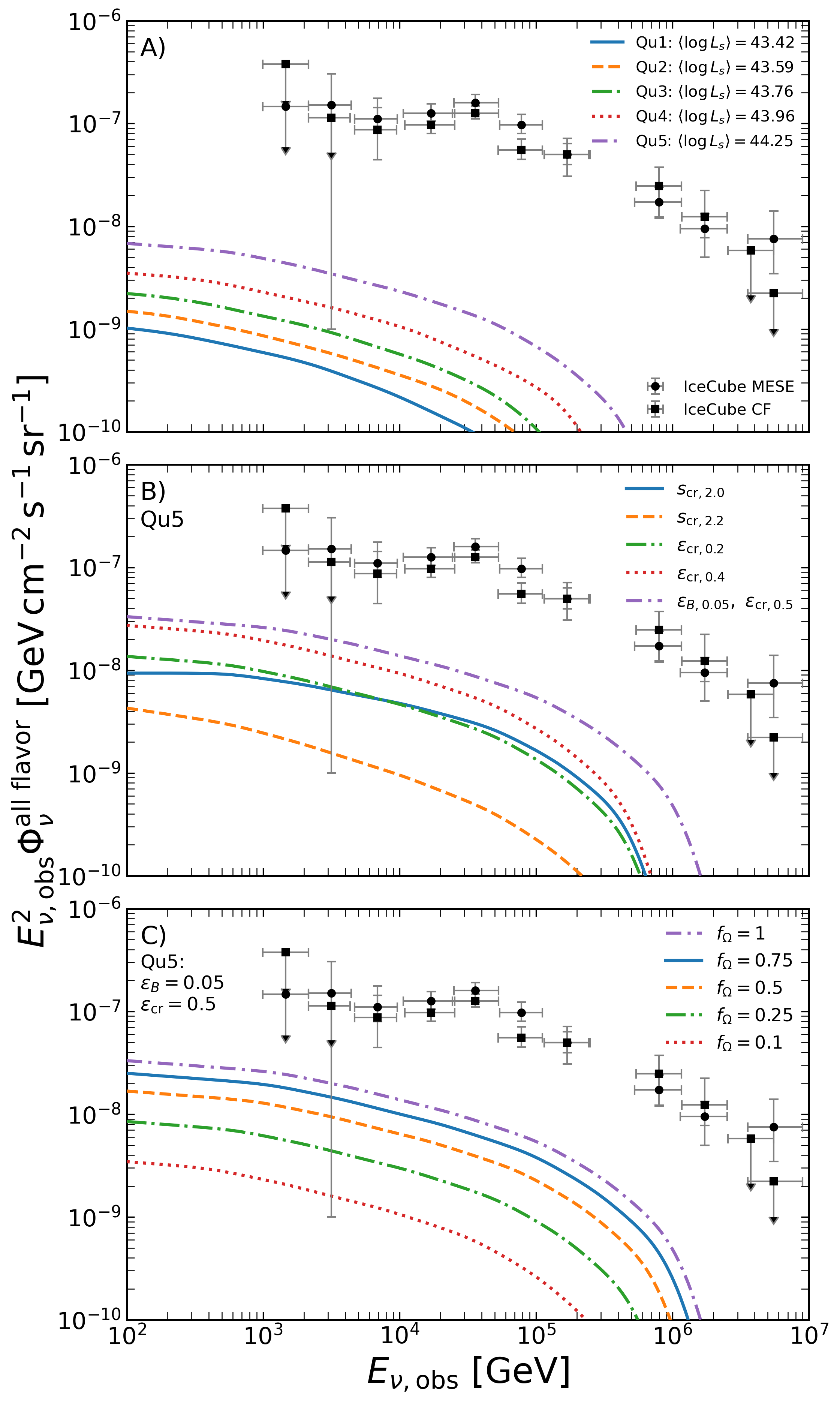}
\caption{A) Diffuse all-flavor neutrino intensity expected from LRDs
compared with IceCube measurements, adopting the fiducial values
$\epsilon_{\rm cr}=0.1$, $\epsilon_B=0.01$, and $s_{\rm cr}=2.1$.
The accepted models shown in Figure~\ref{fig:solutions} are divided
into five equally populated quintiles (Qu1--Qu5) according to their
shock luminosity, $L_s$. Each curve shows the mean diffuse neutrino
intensity of all models within the corresponding quintile.
B) Dependence of the Qu5 neutrino intensity on the assumed cosmic-ray
parameters. We vary $s_{\rm cr}=2.0$ and $2.2$, and
$\epsilon_{\rm cr}=0.2$ and $0.4$, while holding the remaining
parameters the same as A). We additionally show a higher-efficiency
case with $\epsilon_B=0.05$ and $\epsilon_{\rm cr}=0.5$. C) Demonstrating the effect of the covering factor, $f_{\Omega}$, for the higher efficiency case from B). }
    \label{fig:icecube}
\end{figure}

\section{Results}\label{sec:Results}
The bolometric luminosity of shock-powered transient events is typically related to the shock
luminosity through $L_{\rm bol}=\epsilon_{\rm rad}L_s$, where
$\epsilon_{\rm rad}$ is the kinetic-to-radiation conversion efficiency,
with typical inferred values ranging from 0.1 to 0.5 
\citep{2009Fox,2015Fox,2024Brennan,2024WangA}. Based on
\citet{2026Naidu}, we adopt a lower limit on the LRD bolometric
luminosity of $L_{\rm bol}\sim2.0\times10^{43}$~erg~s$^{-1}$. For completeness,
given the uncertain nature of LRDs, we consider the limiting case of
$\epsilon_{\rm rad}=1$, which provides the most conservative lower
bound on $L_s$. Note that lower $\epsilon_{\rm rad}$ would require larger shock luminosities and consequently favor larger neutrino contributions from shock-powered LRDs.
An upper limit is independently provided by the
diffuse neutrino background, whose LRD contribution scales with $L_s$.
From Equation~\ref{eq:diffuse}, requiring that the predicted LRD
contribution does not exceed the diffuse neutrino intensity measured by
IceCube gives $L_s\lesssim3\times10^{45}$~erg~s$^{-1}$ (using
$E_{\rm cr,max}=10^7$~GeV for this constraint).

Guided by physical constraints inferred for LRDs, we explore
$M_{\rm eng}=10^{4}$ to $10^{6}~\Msol$, consistent with the intermediate
central-engine masses proposed for LRDs
\citep{2026Naidu,2026Rusakov, 2026Greene, 2026Umeda, 2026Gentile}. We consider engine mass-loss rates of
$\dot{M}_{\rm eng}=10^{-2}$ to $5~\Msol~{\rm yr}^{-1}$, motivated by
the near- to super-Eddington accretion rates and strong outflows
expected from these systems
\citep{1973Shakura,1999Blandford,2005Ohsuga,2014Jiang,
2024Toyouchi,2025Kido,2026Naidu,2026Kuze}. We vary
$v_{\rm eng}=5,000$ to $15,000~\kms$, motivated by fast central-engine winds in super-Eddington accretion models \citep{2026Kaaz,2026Kuze} and comparable to SN outflows 
interacting with slower surrounding material
\citep{2003Chevalier,2009Dessart,2012Rest,2018Smith}, and test
$n_{\rm H}=10^{9}$ to $10^{12}$~\ccm, representing the dense gas
environments inferred for LRDs
\citep{2025deGraaffb,2026Rusakov,2026Kuze}. We randomly sample these four parameters, calculate $L_s$, and
accept or reject each realization according to the luminosity
constraints and the requirement $M_{\rm esc}\leq M_{\rm eng}$, where
$M_{\rm esc}$ is the total mass escaping the system over the assumed
interaction lifetime.

Figure~\ref{fig:solutions} shows $\approx3000$ accepted parameter sets for each of the slow outflowing and stationary envelope scenarios when $f_{\Omega}=1$. The solutions range from $2.0\times10^{43}\lesssim L_s\lesssim3.5\times10^{44}$~erg~s$^{-1}$.
The hydrogen number density, $n_{\rm H}$, exhibits no strong dependence
within the accepted parameter space. However, for the outflowing envelope models, the sharp drop-off near
$10^{12}$~\ccm\ occurs because the mass-loss rate required to sustain
such high densities exceeds the central-engine mass-loss rate for most
models. For example, using Equation~\ref{eq:density},
$v_{\rm csm}=500~\kms$ requires
$\dot{M}_{\rm csm}\approx1670~\Msol~{\rm yr}^{-1}$ to sustain
$n_{\rm H}=10^{12}$~\ccm. The
central-engine mass similarly exhibits no strong dependence, as it
primarily determines $v_{\rm csm}$, which is not the dominant source
of shock power. In contrast, $v_{\rm eng}$ and $\dot{M}_{\rm eng}$
are strongly constrained, as expected from their direct contribution
to $L_s$. For the outflowing-envelope case, despite the large instantaneous engine mass-loss rates considered, fallback limits the total mass carried outward. The accepted models have a median $M_{\rm esc}/M_{\rm eng}\approx 2.0\times10^{-7}$ over the adopted $40$~yr interaction duration, with $90\%$ remaining below $\sim3.5\times10^{-6}$, indicating that only a negligible fraction of the available engine mass escapes the system.

In the top panel of Figure~\ref{fig:icecube}, we show the potential
contribution of LRDs to the diffuse all-flavor neutrino intensity
compared with IceCube measurements. We sort the accepted
models by $L_s$ and divide them into five equally populated quintiles.
For each model, we calculate the maximum cosmic-ray energy and diffuse
neutrino spectrum individually, with the curves representing the mean
intensity within each quintile. The lower-$L_s$ populations account for only a small fraction of the
observed diffuse intensity, while models in the highest-$L_s$ quintile
contribute on average $\sim2.8\%$ of the IceCube measurements below
$2\times10^4$~GeV and $\sim2.0\%$ below $2\times10^5$~GeV. When
averaged over all IceCube measurements, the contribution is
$\sim1.4\%$. This indicates that the most luminous viable LRD shock
models could provide a small but non-negligible contribution to the
diffuse high-energy neutrino background.

In Panel~B of Figure~\ref{fig:icecube}, we explore the dependence of the
highest-$L_s$ quintile on the cosmic-ray parameters, which may be different for LRDs. A harder cosmic-ray spectrum increases the
relative contribution at high energies. For example, adopting $s_{\rm cr}=2.0$ increases the average
contribution of the highest-$L_s$ quintile to $\sim5.5\%$ below
$2\times10^4$~GeV and $\sim4.2\%$ below $2\times10^5$~GeV.
Increasing $\epsilon_{\rm cr}$ raises the overall neutrino
normalization by increasing the fraction of shock power supplied to
accelerated cosmic rays. Increasing $\epsilon_B$ raises the
magnetic-field strength and therefore $E_{\rm cr,max}$, shifting the
high-energy cutoff of the neutrino spectrum to higher energies. As an
illustrative higher-efficiency case, adopting $\epsilon_{\rm cr}=0.5$
and $\epsilon_B=0.05$ increases the average contribution of the
highest-$L_s$ quintile to $\sim16\%$ below $2\times10^4$~GeV,
$\sim13\%$ below $2\times10^5$~GeV, and $\sim9.5\%$ when averaged over
all IceCube measurements. 

Panel C of Figure~\ref{fig:icecube} illustrates the effects of varying the covering factor, $f_{\Omega}$, for this higher efficiency case to account for potentially asymmetric outflows in LRDs and fragmentation of the stalled shock interaction from radiative instabilities. We consider $f_{\Omega}=1,~0.75,~0.5,~0.25,~{\rm and}~0.1$ and find that decreasing $f_{\Omega}$ reduces the diffuse neutrino contribution proportionally, while shifting the maximum observed neutrino energy to lower values proportional to $\sqrt{f_{\Omega}}$.  We emphasize that Panel~B and C consider only the highest-$L_s$ quintile.
A full population-synthesis study will ultimately be required to
determine the distributions of $L_s$, $f_{\Omega}$, and the cosmic-ray parameters
across the LRD population and thereby predict their total
diffuse neutrino contribution.

These estimates of the neutrino background could turn out to be conservative lower limits with respect to the incompleteness of the observed LRD population.
The current census of LRDs adapted for this calculation is largely limited to so-called ``V-shaped" LRDs \citep[e.g.,][]{2024Kokorev,2025Labbe,2025Kocevski}.
In V-shaped LRDs, the central engine dominates the light from the host galaxy \citep[e.g.,][]{2026Sun}.
However, if the same shock-powered central engine occurred in a bright host galaxy, it would evade V-shaped selection since it is unable to outshine the host galaxy.
Indeed, such examples of buried central engines are now found among gravitationally lensed LRDs, where, if not for lensing, the central engine would be completely anonymous \citep[e.g.,][]{2026Yanagisawa,2026Golubchik,2026Baggen}.
Understanding the true incidence of LRD central engines, beyond V-shaped selection, is still in the early stages \citep[e.g.,][]{2026Weibel,2026Rinaldi,2026Mascia}.
Once these completeness correction factors are in hand, the estimates shown in Figure~\ref{fig:icecube} may increase accordingly.

Conversely, not all observed LRDs may be shock-active. Approximately one-third of a sample of LRD-like sources with high-quality spectra exhibit Balmer absorption or P-Cygni-like features, suggesting the presence of dense gas and, in some cases, slow outflows \citep{2026Rusakov}. These systems may be favorable environments for shock interactions with fast central-engine outflows, although such features do not necessarily establish the presence of shocks. Furthermore, accretion-powered radiation, including X-rays, may be absorbed and reprocessed by dense surrounding gas, contributing to the red optical continuum independently of shock emission \citep{2026Rusakov}. Such accretion-powered emission may coexist with shock interactions or dominate in systems without them. Our calculations assume that all observed LRDs are shock-active, while a smaller fraction would proportionally reduce the predicted diffuse neutrino intensity.

\section{Discussion and Conclusions}  \label{sec:Conclusions}
We have explored whether shock interactions in LRDs can contribute to their observed emission and to the diffuse high-energy neutrino background. We construct a simplified model in which a fast outflow from the central engine initially drives a shock through dense surrounding material before reaching the characteristic LRD photosphere. We consider shocks propagating in both stationary and outflowing surrounding media, followed by a prescribed quasi-stationary interaction near the photosphere. Rather than requiring a single shock to persist over the full LRD lifetime, we adopt an illustrative interaction duration of $\sim40$~yr, with the possibility of successive shock episodes. We find viable solutions over a broad range of central-engine masses and surrounding densities, while the engine mass-loss rate and velocity are more strongly constrained by the required shock luminosity.

The high densities of these models make proton--proton interactions efficient, allowing a fraction of the shock power to emerge as high-energy neutrinos following the framework of \citet{2024Murase}. The resulting neutrino emission depends primarily on the available shock power and the properties of the accelerated cosmic-ray population, allowing the most luminous LRD models to produce an appreciable high-energy neutrino flux.

Finally, we note several limitations of our simplified model. Most importantly, the quasi-stationary interaction is prescribed rather than explicitly modeled hydrodynamically, and we do not explicitly model the mechanism responsible for maintaining the stalled shock. Continued outflows and fallback may facilitate successive interactions, potentially supported by large-scale convective motions in BH-envelope and quasi-star configurations \citep{2026Santarelli,2026Cantiello}, or by fountain flows associated with super-Eddington accretion \citep{2026Kaaz,2026LiuHo}.
Observationally, inverse P-Cygni absorption, indicative of inflowing material, has been reported in $\sim10$--$20\%$ of LRDs \citep{2026Matthee,2026Yanagisawaa,2024Juodvzbalisb,2026Lin,2026Davis}, providing possible evidence for fallback or failed-wind material that could replenish the central regions.
Within this picture, a confined, quasi-stationary shock may form when the outward momentum flux of the central engine is balanced by the inertia and pressure support of the surrounding envelope \citep{1987Rees,2008Begelman,2010Volonteri,2011Ball,2012Ball,2014Baskin,2026Begelman}. Whether such a balance can maintain a stable shock at $R_{\rm ph}$ even over individual interaction episodes ultimately requires multidimensional hydrodynamic calculations.  Furthermore, shock interactions need not account for the entire observed LRD emission. Accretion-powered radiation, including X-rays, may be reprocessed by dense surrounding gas and contribute to the red optical continuum independently of shock emission.

We also adopt values for the cosmic-ray parameters indicative of shock-powered transients. As illustrated in Panel~B of Figure~\ref{fig:icecube}, variations in these quantities can substantially change the predicted neutrino intensity of the highest-$L_s$ models. Panel~C further demonstrates that reducing the covering factor decreases the neutrino intensity and shifts the high-energy cutoff. Better constraints on these parameters, together with a full population-synthesis study of the distribution of $L_s$ across the LRD population, will be needed to establish the total contribution of LRDs to the diffuse neutrino background.

We further fix the stalled-shock location to $R_{\rm ph}=10^{16}$~cm, although the characteristic photospheric radius may vary both between LRDs and within individual sources. BH* models incorporating radial pulsations to explain the long-timescale variability observed in LRDs find photospheric-radius variations of up to several percent \citep{2025Zhang,2026Cantiello}. In our model, a 10\% variation in $R_{\rm ph}$ has little effect on $L_s$, while changing $E_{\rm cr,max}$ by $\sim10\%$ and correspondingly shifting the high-energy cutoff of the neutrino spectrum. Finally, the adopted $n_{\rm LRD}$ may not fully account for LRDs being outshone by brighter host galaxies due to the low completeness of current surveys, potentially increasing the predicted diffuse neutrino intensity. Additionally, not all observed LRDs may be shock-active at a given time. Our calculations assume a shock-active fraction of unity, while a smaller fraction would proportionally reduce the predicted diffuse neutrino intensity.

Overall, our results demonstrate that shock-active LRDs may contribute to the diffuse high-energy neutrino background measured by IceCube. When integrated over the cosmological LRD population, models in the highest-$L_s$ quintile provide an average contribution of $\sim2\%$ of the IceCube diffuse intensity below $2\times10^5$~GeV, decreasing to $\sim1.4\%$ when averaged over all IceCube measurements. For the illustrative higher-efficiency case with $\epsilon_{\rm cr}=0.5$ and $\epsilon_B=0.05$, these contributions increase to $\sim13\%$ and $\sim9.5\%$, respectively. The actual contribution depends on the incidence of shock interactions, the distribution of shock luminosities, and the completeness of the observed LRD population. LRDs may therefore represent an additional population of dense, interaction-powered sources contributing to the diffuse high-energy neutrino background.

\begin{acknowledgments}
T.M., K.M., and C.A. acknowledge support from NASA grants JWST-GO-04522, JWST-GO-04217, JWST-GO-04436,
JWST-GO-03726, JWST-GO-05057, JWST-GO-05290, JWST-GO-06023, JWST-GO-06677, JWST-GO-06213, JWST-GO-06583, JWST-GO-0923. R.P.N. acknowledges support from NASA grants JWST-GO-3516, JWST-GO-5224, and JWST-GO-7404. 
This research is supported in part by JSPS KAKENHI
grant No.~23H04899 (S.H.) and NSF Grant No PHY-2209420 (S.H.). 

{\sl Generative AI disclosure:} ChatGPT (OpenAI) was used under author supervision to assist with manuscript clarity, figure preparation, and code refinement. The authors take full responsibility for the final manuscript and its scientific content.

\end{acknowledgments}



\noindent
{\sl Data Availability Statement:} 
The observational IceCube neutrino data used in this work are publicly available through the \href{https://icecube.wisc.edu/science/data-releases/}{IceCube Collaboration data releases}.
\bibliography{ref}
\bibliographystyle{aasjournalv7}

\end{document}